\documentclass[aps,pra,twocolumn,nofootinbib,showpacs]{revtex4-1}
\usepackage{amsmath}
\usepackage{amssymb}
\usepackage{graphicx}
\usepackage[utf8]{inputenc}
\usepackage[english]{babel}
\usepackage{indentfirst}

\usepackage[usenames]{color}
\usepackage{ulem}
\usepackage[colorlinks=true]{hyperref}

\begin{document}
 
\title{Self-compression of spatially limited laser pulses in a system of coupled light-guides}

\author{A.\,A.\,Balakin}
\email{balakin.alexey@yandex.ru}
\author{A.\,G.\,Litvak}
\author{V.\,A.\,Mironov}
\author{S.\,A.\,Skobelev}
\email{sksa@ufp.appl.sci-nnov.ru}

\affiliation{Institute of applied physics RAS, 603950 Nizhniy Novgorod, Russia}

\date{\today}

\begin{abstract}
The self-action features of wave packets propagating in a two-dimensional system of equidistantly arranged fibers are studied analytically and numerically on the basis of the discrete nonlinear Schr\"odinger equation. Self-consistent equations for the characteristic scales of a Gaussian wave packet are derived on the basis of the variational approach, which are proved numerically for powers $\mathcal{P} < 10 \mathcal{P}_\text{cr}$ exceeding slightly the critical one for self-focusing. At higher powers, the wave beams become filamented, and their amplitude is limited due to nonlinear breaking of the interaction between neighbor light-guides. This make impossible to collect a powerful wave beam into the single light-guide.
The variational analysis show the possibility of adiabatic self-compression of soliton-like laser pulses in the process of their three-dimensional self-focusing to the central light-guide. However, the further increase of the field amplitude during self-compression leads to the longitudinal modulation instability development and formation of a set of light bullets in the central fiber.
In the regime of hollow wave beams, filamentation instability becomes predominant. As a result, it becomes possible to form a set of light bullets in optical fibers located on the ring.
\end{abstract}

\pacs{42.65.-k, 42.50.-p, 42.65.Jx} 
\maketitle
\section{Introduction}\label{sec:1}

The current advance in the technology of generation of ultrashort intense laser pulses not only yields greater pulse intensity and shorter pulse durations, but makes laser facilities less expensive as well, which opens up ample opportunities for a number of important applications \cite{Krausz}. Compression of laser pulses to durations of several optical cycles is based on the spectral-temporal transformation of femtosecond pulses during their propagation through nonlinear media. Extremely short pulses were obtained in nonlinear optical fibers, both in the region of normal dispersion with the subsequent use of dispersion compressors \cite{Fork,Andrianov,Kieu}, and in the region of anomalous dispersion due to multisoliton compression \cite{Sell,Amorim,Foster,Anashkina}. At a millijoule energy level, they were produced in gas-filled capillaries and photon-crystalline fibers due to the combined effect of ionization nonlinearity and plasma dispersion of group velocities \cite{Wagner,Skobelev12,Voronin,Kim15}. A similar mechanism is realized by using filaments \cite{Hauri,Skupin,Stibenz,Kosareva}. 

In the last decade, significant advances have been made in generating high-energy laser pulses in the mid-IR range, for which most media have anomalous dispersion of the group velocity. This has given an impetus to studying radiation self-action in the regime of anomalous dispersion, both in waveguide systems \cite{Balciunas}, and in homogeneous media \cite{Durand,Hemmer}. Here, specific features of the self-action are connected with the fact that radiation self-focusing in a medium with anomalous dispersion is accompanied by shortening of a soliton-like wave packet \cite{Self-focusing}. The possibility of replacing high-power solid-state lasers with equivalent laser systems based on fiber components looks very attractive, since it can change the appeal of the corresponding applied developments in a radical way by ensuring small dimensions, ease of use, reliability, and operation stability of a laser system. 

One of the modern trends in fiber optics is associated with the possibility of using specially structured waveguide systems to control light fluxes \cite{Kivshar, null_1, sth1}. It is actively supported by technological progress in creating micro- and nano-heterogeneous structures. Photonic crystals and metamaterials, in which the optical refractive index is periodically modulated, are widely used \cite{Kivshar, null_1, Sarychev}. At the same time, the nonlinear wave science in spatially periodic media is being developed actively. Due to the nonlinear nature of the interaction of optical radiation with the medium, the situation here turned out to be richer than in solid-state physics. In addition to the purely fundamental interest in the research, there is a practical aspect: the possibilities to generate a supercontinuum \cite{Tran} and shorten the durations of laser pulses \cite{Aceves, Rubenchik}, control the structure of the wave field \cite{null_2, sth2} and form the light bullets \cite{Minardi, Duong}, use active fiber systems for generation of intense laser pulses \cite{null_3}.

In a continuous medium, the basic model for studying the wave field self-action is the nonlinear Schr\"odinger equation (NSE). The generalization of this model for describing the features of nonlinear processes in a system of weakly coupled fibers was successful \cite{Kivshar, null_1, Sarychev, Lederera}. The discrete nonlinear Schr\"odinger equation (DNSE) is also widely used to study the dynamics of nonlinear excitations in solid state physics \cite {Braun} and in molecular systems \cite {Sckot} including those on the quantum level of the description of the medium. Due to the complexity of the processes accompanying the propagation of intense laser radiation in a spatially stratified medium, theoretical research relies mainly on numerical simulation of the problem. It shows that even in the one-dimensional case discrete models demonstrate more complex behavior \cite {Kivshar, null_1, sth1, Sarychev, Lederera, Braun, Sckot} than in the continuous case. The use of approximate methods of studying the system dynamics makes it possible to classify characteristic regimes of self-action and to determine critical parameters. Here, the variational approach has great prospects~\cite {null_4, Trombettoni, Kaup, Skobelev}. 

The purpose of this paper is to study analytically and numerically the possibility of laser pulse self-compression due to the adiabatic decrease in the duration of a soliton-like wave packet during self-focusing of the wave field in a a two-dimensional array of weakly coupled light-guides. Along with that, an important issue is the possibility of coherent summation of radiation into the central light-guide during self-focusing of the radiation in the system under consideration. Based on the analytical and numerical studies performed, we will determine the initial parameters of the laser pulse aiming at achieving the maximum compression of the wave packet in the light bullet mode.

The structure of the paper is as follows. In Section \ref {sec:2}, the system of ordinary differential equations on parameters of the wave packet having the Gaussian form are derived by using the variational approach. On the basis of these equations, a qualitative analysis of the modes of self-action of wave beams in a two-dimensional lattice is carried out, and features of self-compression of femtosecond soliton-like laser pulses in a discrete system are investigated.
Section~\ref{sec:4} presents results of numerical simulation of the two-dimensional DNSE and shows good agreement with analytical analysis of the system dynamics at powers which exceed the critical self-focusing power only slightly.
Using numerical simulation (Section \ref{sec:6}), a detailed picture of the self-focusing of the radiation in the central fiber, shortening and subsequent splitting of the laser pulse with further propagation of the wave field along the axis of the system is investigated. 
The possibility of forming a set of light bullets in the process of development of filamentation instability in the field of hollow wave beams is considered. In the Conclusion, the results of the work are summarized.

\section{Variational approach}\label{sec:2}
Let us consider a model, in which the light-guides are located at the nodes $(n, m)$ of a rectangular lattice. The envelope amplitude $ u_{nm}$ of the wave packet in the $(n,m)$ waveguide changes during the propagation along the $z$ axis under the action of the following factors: dispersion of the waveguide system, cubic nonlinearity of the medium, interaction with neighboring light-guides. This yields the discrete NSE having the following form \cite{Rubenchik, Aceves, Skobelev}:
\begin{multline}\label{eq:1}
i\dfrac{\partial u_{n,m}}{\partial z}+\gamma\dfrac{\partial^2 u_{n,m}}{\partial\tau^2}+ u_{n+1,m}+ u_{n-1,m}+\\+ u_{n,m+1}+ u_{n,m-1}+| u_{n,m}|^2 u_{n,m}=0 .
\end{multline}
Here, $\tau$ is the longitudinal coordinate of the wave packet, and $\gamma$ is the coefficient characterizing the quadratic dispersion of the group velocity of the light-guide. Further, we consider only the case of anomalous dispersion of the group velocity ($\gamma=1$). This equation has a Hamiltonian structure, like the continuous NSE. In addition, system of equations \eqref{eq:1} preserves the total energy of the wave packet
\begin{equation}\label{eq:2}
W = \int\limits_{-\infty}^{+\infty}\sum\limits_{n,m}| u_{n,m}|^2d\tau=\text{const}.
\end{equation}
In the case of continuous radiation $\partial_{\tau\tau}  u_{nm}=0$, the conserved quantity is the power
\begin{equation}\label{eq:3}
\mathcal{P}=\sum\limits_{n,m}| u_{n,m}|^2 .
\end{equation}

To study the peculiarities of the discrete problem qualitatively, we turn to the variational approach and then compare the results of the approximate analysis with the numerical solution of system \eqref{eq:1}. As in the NSE~case, the variational approach allows one to obtain ordinary differential equations for changing the width of the wave field and the curvature of the phase front of Gaussian wave beams in a discrete problem along the propagation path. As a result, it is possible to describe not only the self-focusing of initially smooth field distributions up to the lattice size, but also the effects of the transition to the self-channeling mode of radiation in the central fiber \cite {Skobelev}, which are determined by the discrete nature of the medium. This special case of the variational approach based on the Gaussian form for wave pulses is often called the \emph{aberration-free approximation}. This approach was firstly introduced in \cite{AF_first} and is widely used in nonlinear optics for continuous media \cite{Self-focusing}.

To obtain analytical results, we use the approach based on the Lagrangian
\begin{multline}\label{eq:4}
\mathcal{L}=\sum\limits_{n,m=-\infty}^{+\infty}\dfrac{i}{2} \left( u_{n,m}\dfrac{\partial u_{n,m}^\star}{\partial z}-c.c. \right) + \left|\frac{\partial  u_{nm}}{\partial \tau}\right|^2\\
-\left( u_{n+1,m} u_{n,m}^{\star}+ u_{n,m+1} u_{n,m}^{\star}+c.c. \right)-\dfrac12| u_{n,m}|^4.
\end{multline}
By using the Poisson summation formula
\begin{equation}\label{eq:5}
\sum\mathcal{F}(n,m,z)=\int \mathcal{F}(x,y,z) \sum e^{2\pi inx+2\pi imy}dxdy ,
\end{equation}
expression \eqref{eq:4} is converted to the form
\begin{multline}\label{eq:6}
\mathcal{L}=\sum\int\Big[\dfrac{i}{2}\left( u\dfrac{\partial u^{\star}}{\partial z}-c.c. \right) + \left|\frac{\partial  u_{nm}}{\partial \tau}\right|^2 \\ 
-\big( u(x+1,y,z) u^{\star}(x,y,z)+ u(x,y+1,z) u^{\star}(x,y,z) +\\ c.c. \big)  -\dfrac12| u|^4 \Big]e^{i2\pi(nx+my)}dx dy,
\end{multline}
which allows us to describe the evolution of a discrete system by a function of the continuous argument $ u(x, y, z)$. As a result, the problem is reduced to a ``continuous'' one, and for analysis of this problem, it is natural to use the variational approach.

Next, we will study the evolution of axially symmetric Gaussian wave packets 
\begin{multline}\label{eq:7}
 u_{n,m}=\dfrac{\sqrt{W}}{a\sqrt{\tau_0}\sqrt[4]{\pi^3}} \exp\Big(-\dfrac{\tau^2}{2\tau_0^2}-\dfrac{(x^2+y^2)}{2a^2}+\\ \dfrac{}{}+i\alpha(x^2+y^2)+i\beta\tau^2 \Big).
\end{multline}
The parameters $\tau_0$ and $\beta$ characterize the duration and frequency modulation (chirp) of the wave packet. The parameters $a(z)$ and $\alpha(z)$ describe the change in the beam width and the curvature of its phase front during propagation.

In the considered case of the Gaussian field distributions (Eq.\,\eqref{eq:7}), we can perform integration in \eqref{eq:6} and obtain an expression for the Lagrangian in the form of a functional series. Estimates for the series terms show that it is sufficient to take into account the single term with $n=m=0$ to describe the processes with the width $a \gg 1/\pi$. This condition actually means that the proposed approximation of the discrete field distribution by continuous function \eqref{eq:7} remains valid also for describing the evolution of distributions with a characteristic scale comparable with the lattice size. As a result, we arrive at the following truncated Lagrangian of the system
\begin{multline}\label{eq:8}
\mathcal{L}_0 = \dot{\alpha} a^2 W + \frac{1}{2} \dot{\beta} \tau_0^2 W + \frac{W}{2 \tau_0^2} \left( 1+4\beta^2 \tau_0^4 \right) -\\ - 4W\exp\left(-\dfrac1{4a^2}-\alpha^2a^2\right) - \dfrac{W^2}{\sigma a^2 \tau_0},
\end{multline}
where $\sigma = 4 \pi \sqrt{2 \pi}$, $\dot{q}\equiv dq/dz$, and $q$ are the parameters $\{a, \tau_0,\alpha,\beta\}$. Using the Euler equation
$$\frac{d}{dz} \frac{\partial\mathcal{L}_0}{\partial \dot{q}} - \frac{\partial\mathcal{L}_0}{\partial q} = 0, $$
we obtain
\begin{subequations}\label{eq:10}
	\begin{gather}
	\dot{\alpha} = \left(\dfrac1{a^4}-4\alpha^2\right) e^{ -\frac{1}{4a^2}-\alpha^2a^2} - \frac{W}{\sigma a^4 \tau_0}, \label{eq:10a} \\
	\dot{a} = 4\alpha a e^{ -\frac{1}{4a^2}-\alpha^2a^2}, \label{eq:10b} \\
	\ddot{\tau}_0 = \frac{4}{\tau_0^3} - \frac{4}{\sigma} \frac{W}{a^2 \tau_0^2}. \label{eq:10c}
	\end{gather}
\end{subequations}
Equations \eqref{eq:10a} and \eqref{eq:10c} describe the competition of diffraction and nonlinear refraction (first and second terms, respectively). The nonlinear alteration of the phase front is determined by the same expression as that in a continuous medium. The main contribution made by media discreteness is the exponential weakening of the diffraction effects.

Note that equations \eqref{eq:10} have a stationary solution with $\alpha=0$, $\dot{a}=0$, $\dot{\tau}_0=0$ and
\begin{equation}\label{eq:11}
\mathcal{P}\equiv \frac{W}{\sqrt{2\pi} \tau_0} = 4\pi\exp\left(-\frac{1}{4a_s^2} \right),
\end{equation}
where $a_s$ is the width of the homogeneous waveguide channel. There are no stationary solutions for $\mathcal{P}>\mathcal{P}_\text{cr} = 4\pi$. Exactly this regime of self-action will be considered below. For $\mathcal{P}\le 4\pi$, the stationary solutions correspond to the discrete analogue of the Townes mode in a continuous media \cite{Tawnes64}. However, these solutions are unstable according to the generalized Vakhitov-Kolokolov criteria~\cite{Laedke95}, similarly to those in the continuous case. 

\subsection{Collapse of 2D wave beams}

Let us first consider the stationary wave beam dynamics (i.e., assume that $\tau_0 \to \infty$). 
In this case, system of equations \eqref{eq:10a}, \eqref{eq:10b} have the integral
\begin{equation}\label{eq:12}
\exp\left(-\dfrac1{4a^2}-\alpha^2a^2 \right)+\dfrac{\mathcal P}{16\pi a^2}=\mathcal{C} .
\end{equation}
In the case under consideration, $\mathcal C$ is a quantity proportional to the Hamiltonian. From Eq.~\eqref{eq:12}, we can find the expression for the curvature of the phase front
\begin{equation}\label{eq:13}
\alpha^2=\dfrac1{a^2}\left[-\ln\left(\mathcal{C}-\dfrac{\mathcal P}{16\pi a^2} \right)-\dfrac1{4a^2} \right] .
\end{equation}
This expression is meaningful only for wave beams having a size greater than the minimal one:
\begin{equation}\label{eq:14}
a_\text{min}=\sqrt{\dfrac{\mathcal P}{16\pi\mathcal{C}} }.
\end{equation}
In the case of an initially wide collimated wave beam ($a_0 \gg a_\text{min}$, $\alpha_0 = 0$), the integration constant is $\mathcal{C} \simeq 1$. It is easy to see that the right-hand side of \eqref{eq:13} is positive for the power $\mathcal{P}> 4\pi$. This corresponds to the self-focusing condition in the continuous case.

Excluding $\alpha$ from \eqref{eq:10b}, we obtain the equation
\begin{equation}\label{eq:15}
\dfrac{da}{dz}=\pm 4 \Big(\mathcal{C}-\frac{\mathcal P}{16\pi a^2} \Big) \sqrt{-\ln \Big(\mathcal{C}-\frac{\mathcal P}{16\pi a^2} \Big) - \frac{1}{4a^2}} ,
\end{equation}
describing the evolution of the beam width. Its right-hand side contains two cofactors. The spatial dynamics of broad wave beams ($a \gg a_\text{min}$) with a power exceeding the critical one is determined by the second cofactor. The qualitative difference from a continuous medium arises with decreasing of beam width and is associated with the role of the first cofactor in Eq.~\eqref{eq:15}. Thus, we can distinguish two evolution stages of wide wave beams ($\mathcal{C} \approx 1$). At the first one (while $\mathcal{P} \ll 16 \pi a^2$), the radiation self-focusing occurs as in the continuous case. It is described by the equation
\begin{equation}\label{eq:16}
\dfrac{da}{dz}=-\dfrac1{\sqrt{\pi}a}\sqrt{\mathcal{P}-4\pi} .
\end{equation}
Hence, for the self-focusing length, we find
\begin{equation}\label{eq:17}
L_\text{sf}=\dfrac{a_0^2\sqrt{\pi}}{2\sqrt{\mathcal{P}-4\pi}} .
\end{equation}
During the self-focusing of the beam, the discreteness of the medium begins to count, and the system goes into a mode, in which the first cofactor in Eq.~\eqref{eq:15} vanishes. 

At this final stage, the evolution of the width is determined by the equation
\begin{equation}\label{eq:18}
\dfrac{da}{dz}=-4 \Big(\mathcal{C}-\dfrac{\mathcal P}{16\pi a^2}\Big) \sqrt{-\ln \Big(\mathcal{C}-\dfrac{\mathcal P}{16\pi a^2} \Big)} .
\end{equation}
It describes the decrease in the wave beam width to the minimum size \eqref{eq:14} by the asymptotic law at $z\to \infty$
\begin{equation}\label{eq:19}
a \approx \sqrt{\frac{\mathcal P}{16\pi}} \left[1+\frac{1}{2} \exp \Big(-\dfrac{256 \pi}{\mathcal P} z^2 \Big) \right] .
\end{equation}
Hence, the characteristic length of formation of a homogeneous waveguide channel is much smaller than the self-focusing length, Eq.\,\eqref{eq:17}. The smallness of this length justifies the separation of two stages in the field evolution: self-focusing and formation of a ``homogeneous'' waveguide structure. It is important to note that the self-channeling mode differs significantly from the continuous one. The phase front of the wave beam in this regime is not flat. From \eqref{eq:10a} we see that the wave front curvature increases linearly with respect to $z$ along the propagation path, as soon as the width decreases to a minimum size. Thus, this process corresponds more likely to a collapse to the wave beam structure having a finite width.

\subsection{Self-compression of 3D wave packets}\label{sec:5}

Let us now turn to studying the features of self-compression of laser pulses in a discrete system. In this case, it is necessary to analyze the complete, four-dimensional system of equations \eqref{eq:10}. However, for sufficiently short pulses $\tau_0 \ll a$, we can select a slow-motion trajectory in system \eqref{eq:10}, which corresponds to soliton-like field distributions along the longitudinal coordinate. For such distributions, the right-hand side of \eqref{eq:10c} is close to zero. This means that the pulse duration changes smoothly during self-focusing of the radiation according to the law
\begin{equation}\label{eq:23}
\tau_0(z) \approx \frac{\sigma}{W} a^2(z).
\end{equation} 
Excluding $\tau_0$ from \eqref{eq:10a}, we obtain
\begin{equation}\label{eq:24}
\dot{\alpha} = \left(\dfrac1{a^4}-4\alpha^2\right) e^{ -\frac{1}{4a^2}-\alpha^2a^2} - \frac{W^2}{\sigma^2 a^6}.
\end{equation}

So, the self-action of soliton-like pulses is described by equations \eqref{eq:10b} and \eqref{eq:24} having an integral similar to \eqref{eq:12}:
\begin{equation}\label{eq:26}
\exp\left(-\dfrac1{4a^2}-\alpha^2a^2\right)+\frac{W^2}{2 \sigma^2 a^4}=\mathcal{C} .
\end{equation}
As above, the integration constant $\mathcal{C}\simeq 1$ for a wide initial collimated wave packet ($a_0 \gg 1$, $\alpha_0=0$).

Excluding $\alpha$ from \eqref{eq:26} we found the following equation for the wave beam width
\begin{equation}\label{eq:27}
\dfrac{da}{dz}=\pm 4 \Big(1-\dfrac{W^2}{2 \sigma^2 a^4} \Big) \sqrt{-\ln \Big(1-\frac{W^2}{2 \sigma^2 a^4} \Big) - \frac{1}{4a^2}}.
\end{equation}
The situation here is somewhat more complicated than for analogous equation \eqref{eq:15} in the two-dimensional case. As above, the evolution of a wide wave packet ($a^2 \gg W/\sigma$) proceeds as in a continuous media and is determined by the second cofactor in Eq.~\eqref{eq:27}. However, the critical energy for self-focusing in the three-dimensional case depends on the initial width of the wave beam $a_0$:
\begin{equation}\label{eq:28}
W>W_\text{cr} = \sqrt{2} \sigma a_0.
\end{equation}
If this relation is satisfied, then the three-dimensional collapse takes place (both the beam width and its duration decreases according to \eqref{eq:23}). In the process of self-focusing, the role of the first cofactor in \eqref{eq:27}, which is determined by the media discreteness, increases. As a result, the asymptotic law at $z\to\infty$ is described by the equation
$$ \frac{da}{dz}=-4 \Big(1-\frac{W^2}{2 \sigma^2 a^4}\Big) \sqrt{-\ln \Big(1-\frac{W^2}{2 \sigma^2 a^4}\Big)}. $$
Its solution
\begin{equation}\label{eq:31}
a=a_c \Big[1+\frac{1}{4} e^{-{64z^2}/{a_c^2}} \Big], \quad a_c=\sqrt{W/\sigma}.
\end{equation}
describes the regime of radiation self-focusing up to the characteristic size $a_c$. In this case, according to \eqref{eq:23}, an adiabatic decrease in the pulse duration up to the value takes place
\begin{equation}\label{eq:32}
\tau_c=1.
\end{equation}

Thus, the qualitative study shows that the radiation self-focusing is accompanied by a noticeable shortening of the  duration of three-dimensional wave packets with the soliton-like distribution along the longitudinal coordinate. At the initial stage, the self-action process develops as in the continuous media and is determined by the same condition, Eq.\,\eqref{eq:28} \cite{Skobelev16}. The features of the discrete medium become apparent at the final stage and lead to the finite size of the field localization in the transverse and longitudinal directions (\eqref{eq:31} and \eqref{eq:32}, respectively). As a result, a localized structure, called \emph{a light bullet}, is formed.

\section{Results of numerical simulation of wave beam self-focusing}\label{sec:4}

\begin{figure*}
	\includegraphics[width=0.77\linewidth]{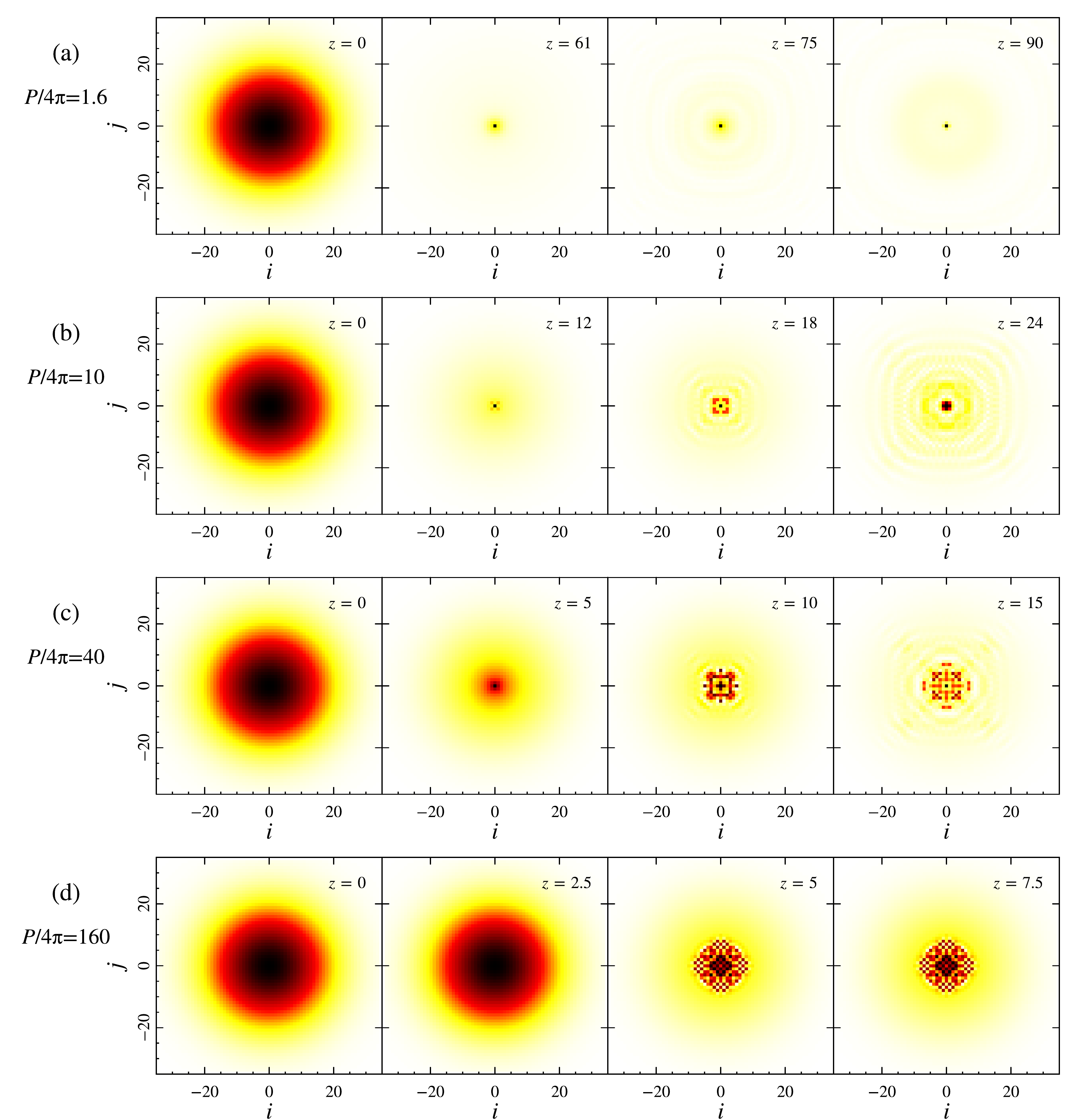}
	\caption{(Color online) Evolution of the amplitude distribution of a wave beam during its self-focusing in a discrete medium. The initial field distribution is $u_{ij}=\sqrt{P/40} e^{-(i^2+j^2)/{320}}$. }\label{ris:ris1}
\end{figure*}

\begin{figure*}
	\includegraphics[width=0.8 \linewidth]{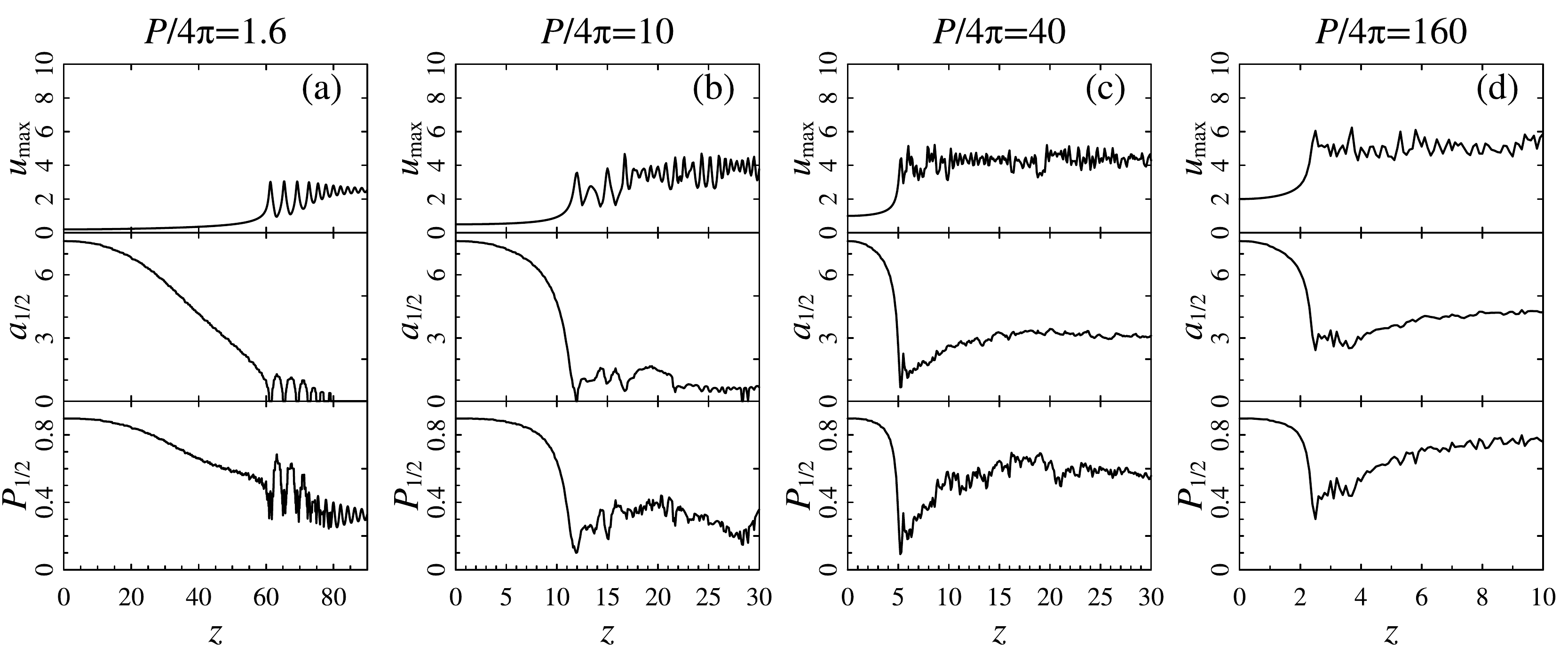}
	\caption{Maximal amplitude $u_\text{max}$, the width $a_\text{1/2}$ and the power fraction $\eta_\text{1/2}$ in fibers with high intensity depending on the evolution variable $z$ for different values of the initial power $\mathcal{P}$. The limitation of the maximum amplitude, associated with the development of the filamentation instability, is seen. The dashed lines present $u_\text{max}=\sqrt{2\pi}$. The parameters are identical with those in Fig.~\ref{ris:ris1}. }\label{ris:ris2}
\end{figure*}

\begin{figure}
	\includegraphics[width=\linewidth]{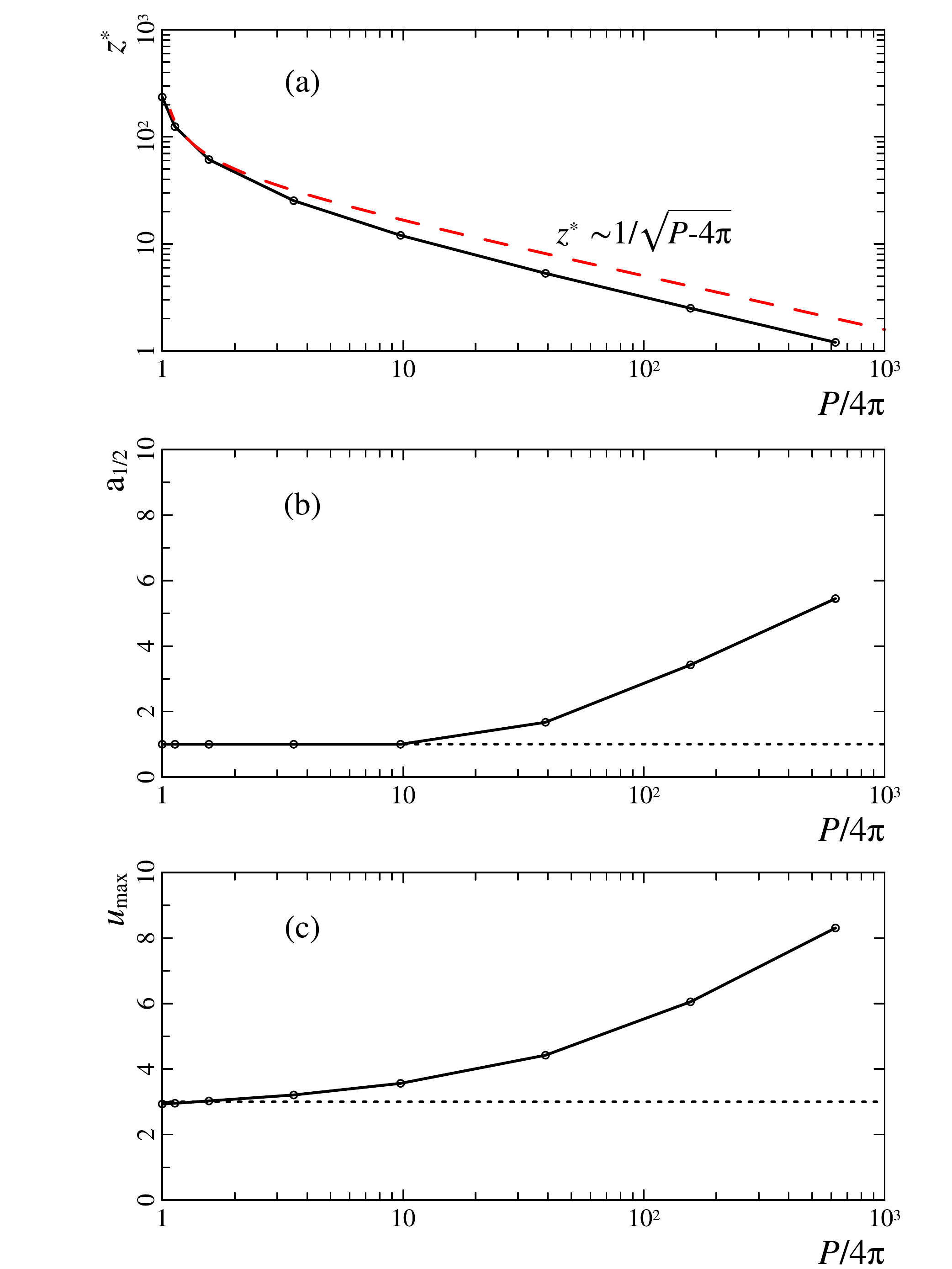}
	\caption{\textbf{(a)} Dependence of the length of self-focusing on the wave beam power. The dashed curve corresponds to analytical expression \eqref{eq:17}, and the solid one is the result of numerical simulation. Figure \textbf{(b)} shows the dependence of the effective localization region $a_{1/2}$ and its estimate \eqref{eq:est_a12} on the power $\mathcal{P}$, the root-mean-square width $\langle a\rangle$ and the estimate for minimal width \eqref{eq:14}, correspondingly. \textbf{(c)} Dependence of the maximal output amplitude $u_\text{max}$ on the power $\mathcal{P}$. }\label{ris:ris3}
\end{figure}

Numerical simulation of initial equations \eqref{eq:1} in media without dispersion shows that the variational approach describes quite well the features of the discrete problem for a power that is of the order of  the critical value $\mathcal{P}_\text{cr} = 4\pi$. The data presented in figures \ref{ris:ris1}~--~\ref{ris:ris3} give a complete picture of the evolution of the initially wide (on the lattice scale) Gaussian wave beam \eqref{eq:7} with $\alpha_0=0$. At a power of $\mathcal{P}$ exceeding the critical self-focusing power $\mathcal{P}_\text{cr}$, the radiation is focused and the wave beam is self-channeled near the system axis (see Fig.~\ref{ris:ris1}). 
Data processing shows (see Figures~\ref{ris:ris1}\textbf{(a)}, \ref{ris:ris2}\textbf{(a)}, and \ref{ris:ris3}) that the radiation becomes localized in the central fiber for $\mathcal{P} < 10 \mathcal{P}_\text{cr}$. The field profile here is close to the Gaussian form. Thus, both the root-mean-square width $\langle a\rangle =\sqrt{\sum_{ij} (i^2+j^2)|u_{ij}|^2/P}$ and the minimal width estimate \eqref{eq:14} coincide rather well (Fig.~\ref{ris:ris3}\textbf{(b)}). Correspondingly, the decrease in the self-focusing length (Fig.~\ref{ris:ris3}\textbf{(a)}) takes place according to the results of the variational-approach analysis (Eq.~\eqref{eq:16}). At this, the maximal field amplitude is limited by a value of about 3. 

A further increase in the power of the wave beam, $\mathcal{P} \ge 10 \mathcal{P}_\text{cr}$, is accompanied by an increase in the effective region of self-localization of the field. With increasing radiation power, one can see the development of the filamentation instability and the wave beam splitting (see Fig.~\ref{ris:ris1}\textbf{(b)}--\textbf{(d)}). In continuous media with the Gaussian field distribution, this process was studied in~\cite{Fibich}. It was shown that wave beam splitting develops if the power is of an order of magnitude exceeding that of the critical power of self-focusing. 

Another noticeable difference between the discrete problem and the continuous one is associated with the diffusion of the wave field into the peripheral region during the wave-beam self-focusing. In the one-dimensional case, this effect is described in \cite{Skobelev}. Its magnitude can be estimated on the basis of Fig.~\ref{ris:ris2}, which shows the change in the power fraction $\eta_{1/2}$ in the central part of the wave beam (at $1/2$ level) along the propagation path. Numerical simulations give about half of the initial power captured in the self-trapping mode.

At a power exceeding $\mathcal{P}_\text{cr}$ by orders of magnitude, a typical aureole appears (see Fig.~\ref{ris:ris1}), which has a characteristic size about the initial wave beam width. This aureole focuses on the center region even after the collapse. This results in 
an increase in both the power fraction $\eta_{1/2}$ in the axial part and the characteristic size $a_{1/2}$ of the intense-field region with an increasing wave beam power (see Fig.~\ref{ris:ris2}\textbf{(c)}). The noticeable difference between the root-mean-square width $\langle a\rangle$ and width $a_{1/2}$ denotes the presence of a large aureole for this case. At this, the variational approach and estimate \eqref{eq:14} become inapplicable due to appearance of several local maximums at $a_{1/2}>0$.

For a discrete system, the necessary conclusions can be drawn from the expression for the growth rate of the $\Gamma$ filamentation instability of a plane wave with the amplitude $ u_0$. The corresponding $\Gamma$ dependence of the perturbation $\propto \exp (i \kappa n) $ on the wave number $\kappa$ has the form \cite{Braun}
\begin{equation}\label{eq:Gamma}
\Gamma^2=4\sin^2\frac{\kappa}{2}\left(2 u_0^2-4\sin^2\frac{\kappa}{2} \right) .
\end{equation}
It follows that the instability growth rate for $ u_0^2>2$ is maximal at the transverse scale of the field perturbation equal to the lattice period
\begin{equation}\label{eq:Lperp}
L_{\perp} \equiv \pi/\kappa =1 \quad\text{for}\quad u_0 > \sqrt{2}.
\end{equation}

The numerical simulation demonstrates splitting of self-channeled radiation with the fields amplitudes $ u_0 \gg \sqrt{2}$. Thus, at the nonlinear stage the development of filamentation instability in a discrete problem leads to the wave beam splitting into a set of wave structures localized in separate optical fibers. Such sets of soliton-type distributions were investigated in papers \cite{Mezencev, Efremidis}.

The fields in each fiber will have their own nonlinear wave-number shift $h \simeq | u_{n,m}|^2$, according to \eqref{eq:1}. The interaction between the light guides will be exponentially weak if the phase difference $L_d (h - h')$ for the characteristic ``diffraction'' length of one lattice cell $L_d \simeq 1/2$ is much larger than $\pi$. At the nonlinear stage, this limits the growth of the field amplitude by the magnitude of the order of $u_\text{max}$:
\begin{equation}\label{eq:umax}
L_d (u_\text{max}^2 - |u|^2) \gg \pi ~ \Rightarrow ~ u_\text{max} \gtrsim \sqrt{2\pi + u_0^2} \mathop{\approx} \limits_{u_0 \ll 1} \sqrt{2\pi},
\end{equation}
where $u_0$ is the background (initial) amplitude.
Numerical simulation shows that the output amplitude lies in the range $\sqrt{2\pi} \ldots 3 \sqrt{2\pi}$, when the power $\mathcal{P}$ changes by 3 orders of magnitude (Fig.~\ref{ris:ris2}, \ref{ris:ris3}\textbf{(c)}). At these amplitudes, the nonlinear influence on the field evolution becomes stronger than the tunneling of the wave field into neighboring fibers, and self-channeling takes place in a narrow set of fibers.

The limitation of the maximal amplitude allows us to estimate the number of inhomogeneities $\mathcal{N} \approx \eta_{1/2} P/u_\text{max}^2$.
Indeed, using Eq.~\eqref{eq:umax} for the maximal amplitude and assuming $\eta_{1/2} \approx 0.5$, one should expect about 20 and 80 inhomogeneities for powers in Figs.~\ref{ris:ris1}\textbf{(c)} and \ref{ris:ris1}\textbf{(d)}, correspondingly.
The width of the region with a strong field can be estimated as
\begin{equation}\label{eq:est_a12}
a_{1/2} \approx \frac{\mathcal{N}}{2} \approx \frac{1}{2} \sqrt{\frac{\eta_{1/2} P}{u_\text{max}^2}} \approx \frac{1}{8} \sqrt{\frac{P}{\pi}}.
\end{equation}
Here, we use $u_{max} \approx 2\sqrt{2\pi}$, $\eta_{1/2} \approx 0.5$ for the final estimate. Numerical simulations show good agreement with this estimate (Fig.~\ref{ris:ris3}\textbf{(b)}).

\begin{figure}
	\includegraphics[width=0.9 \linewidth]{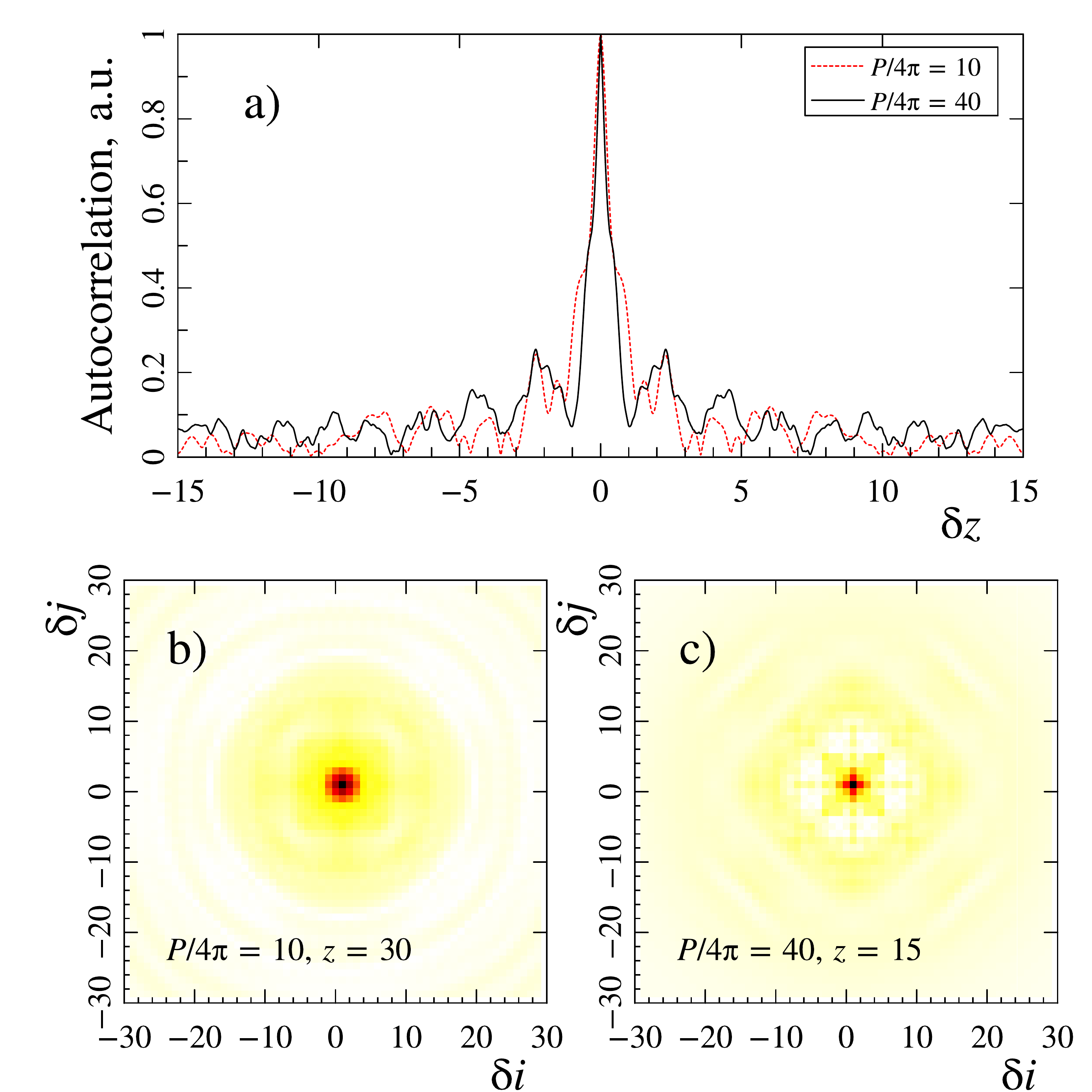}
	\caption{(Color online) Autocorrelation function for the wave field envelope along the evolutionary coordinate in the central fiber \textbf{(a)} and along the transverse coordinates \textbf{(b, c)} for different power values.}\label{ris:ext}
\end{figure}

Along with the filamentation in the central part, the focusing continues at the periphery of the wave beam. This is responsible for a certain expansion of the region occupied by the intense field $a_{1/2}$ immediately after the collapse point.
Moreover, the presence of the instability only in a localized region leads to the rise of a complex, stochastic field dynamic for powers significantly exceeding the critical one. We calculate the autocorrelation functions for the modification of the stratified structure during propagation of the wave field $K_{ij}(\delta z) = \int \big(u_{ij}(z) u_{ij}^*(z+\delta z)+\text{c.c.}\big) dz$, and for the presence of a strongly localized maximum $K_\perp(\delta i, \delta j) = \sum \big(u_{ij} u_{i+\delta i,j+\delta j}^*+\text{c.c.}\big)$ at $\delta z=0$ to testify this. The first autocorrelation function characterizes the temporal coherence, the second one characterizes spatial coherence. Panels ({\em  a}--{\em  c}) in Fig.~\ref{ris:ext} show the strongly localized maxima at $\delta z=0$ and $\delta i=\delta j=0$, which clearly manifests the stochastic dynamics.

Note that the discrete collapse will not lead to a collapse in a continuous media in general. Indeed, the maximal power of the field in dimensional units is determined by the ratio of the coupling coefficient between adjacent light-guides and the nonlinear coefficient. For the case~\cite{Rubenchik}, the coupling coefficient is about 0.01~cm$^{-1}$ and the nonlinear coefficient is about $10^{-5}$~(W\,cm)$^{-1}$. At this, the maximal amplitude \eqref{eq:umax} in dimensionless units is limited by $|u_\text{max}|^2 \lesssim 2 \pi$. Altogether, they give the maximal power in the single light-guide of the order of
$$ \frac{P_\text{max}}{P_\text{cr}} = \frac{|u_\text{max}|^2}{9 \times 10^6 \,\text{W}} \frac{0.01 \,[\text{cm}^{-1}]}{10^{-5}\, [(\text{W}\,\text{cm})^{-1}]} < 0.01. $$
So, the maximal value is much smaller than the critical power $P_\text{cr}$ for self-focusing, which is about 9~MW for fused silica. As a result, the single-mode approach used for Eq.~\eqref{eq:1} is still valid, and the effects of nonlinearities inside the light-guides are weak. The similar estimate was proved by solving the original NSE \cite{Eilenberger} for large coupling coefficients (of about 1~cm$^{-1}$).

Finally, we've shown that a two-dimensional array of light-guides is not suitable for collecting the power to a single light-guide. The issue is the limitation of maximal amplitude \eqref{eq:umax}, which effectively breaks the interaction between neighboring light-guides for higher amplitudes. This effectively limits oneself by using radiation with powers essentially above the critical one. Fortunately, this problem can be bypassed for pulsed radiation, if we take into account media dispersion, which opens two new possibilities. The first one is ejecting the pulse energy from the central part by pulse splitting. The second is formation of a high-power narrow pulse in the central light-guide, whose amplitude will increase due to the energy fluxes at tails from neighboring light-guides. The both cases can yield the pulse amplitude above the threshold (Eq.\,\eqref{eq:umax}). These possibilities will be considered in the next section.

\section{Light bullet formation}\label{sec:6}

The collapse of three-dimensional wave packets, even within the framework of continuous NSE, is one of those insufficiently investigated processes \cite{Self-focusing}, especially in the case of wave packets, which are oblate in longitudinal direction \cite{Skobelev16}. The specificity of the problem we are considering is that the media discreteness manifest itself at the wave field self-focusing, and it becomes possible to capture radiation into the self-channeling regime (see section~\ref{sec:5}).

\begin{figure}
	\includegraphics[width=0.8\linewidth]{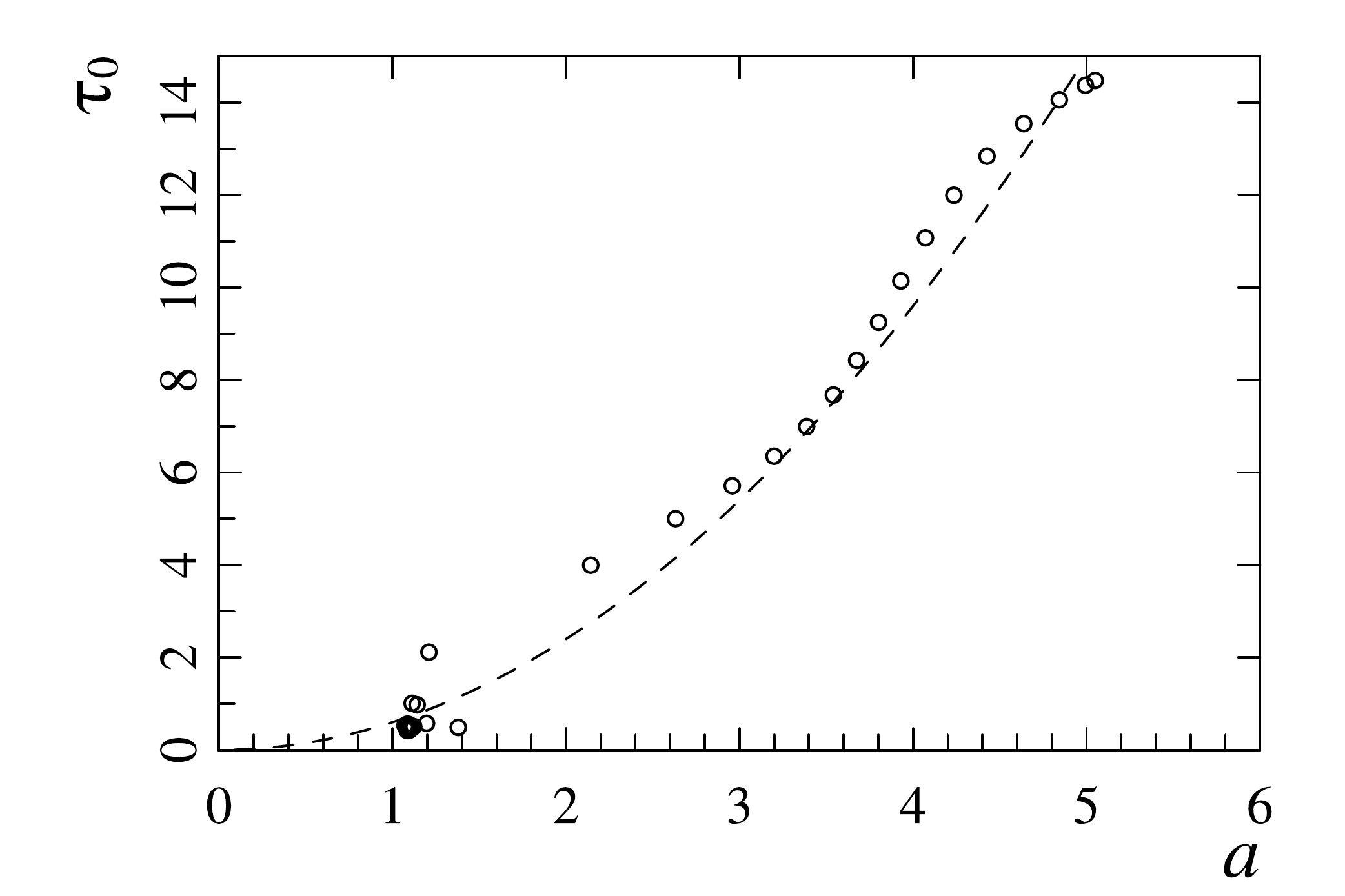}
	\caption{Dependence of the wave packet duration $\tau_0$ on its width $a$ during the self-focusing process. The dashed curve is described by formula \eqref{eq:23} for $W = 157$. A curve consisting of hollow circles is determined on the basis of numerical simulation data of original equation \eqref{eq:1}. The initial wave packet form is $u_{nm}=0.4 e^{-\left((n-25)^2+(m-25)^2 \right)/{50} - {\tau^2}/{100}}$. }\label{ris:ris4}
\end{figure}

\subsection{Gaussian wave pulses}

\begin{figure*}
	\includegraphics[width=0.75\linewidth]{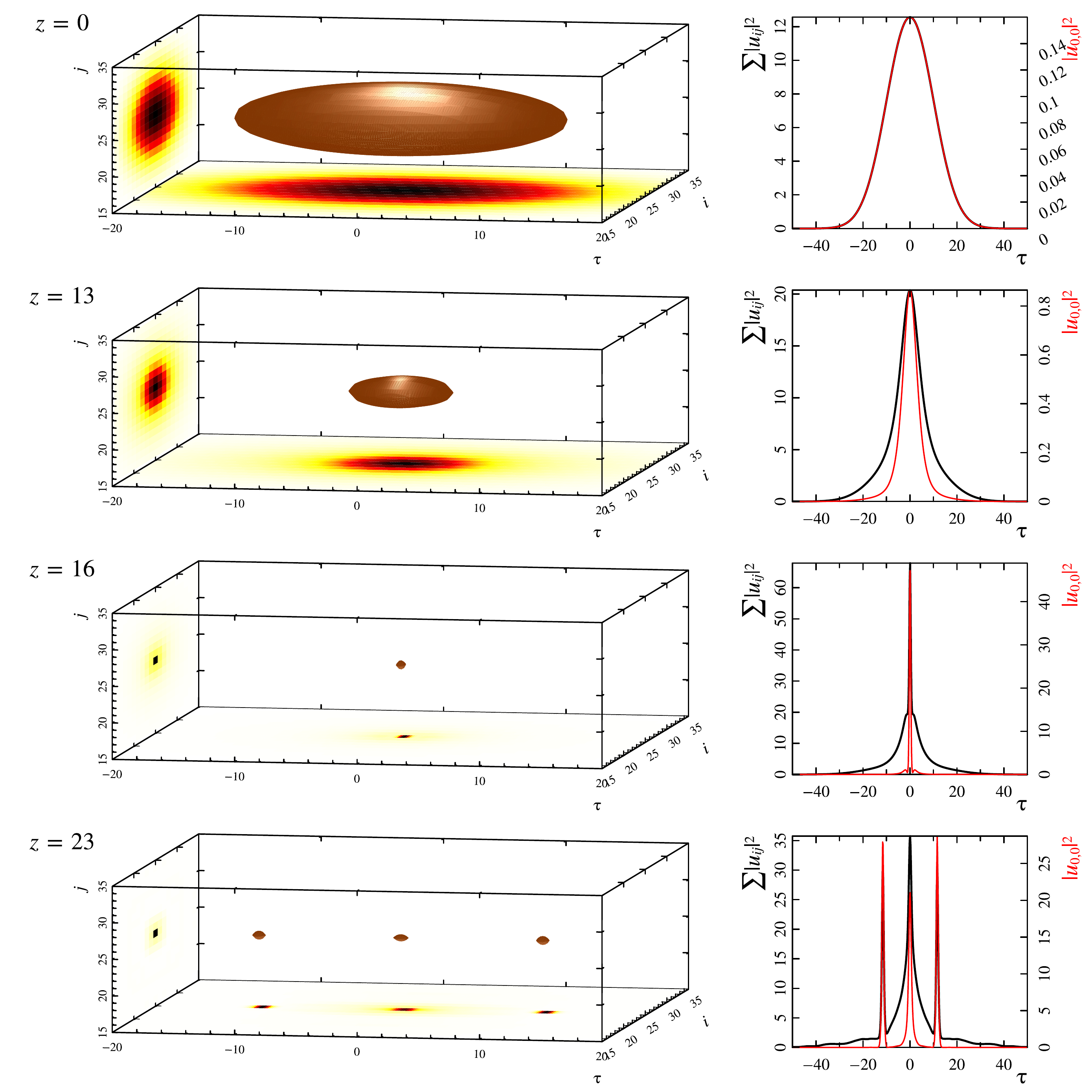}
	\caption{(Color online) The left column shows the amplitude distributions of the wave packet with the energy exceeding the critical one at different values of $z$. The right column shows the corresponding time structures of the wave packet: integral one (black) and one in the region of the maximum field (red). The initial pulse is $u_{ij}=0.4 e^{-\left((i-25)^2+(j-25)^2 \right)/{50} - {\tau^2}/{100}}$. }\label{ris:ris5}
\end{figure*}

\begin{figure*}
	\includegraphics[width=0.9 \linewidth]{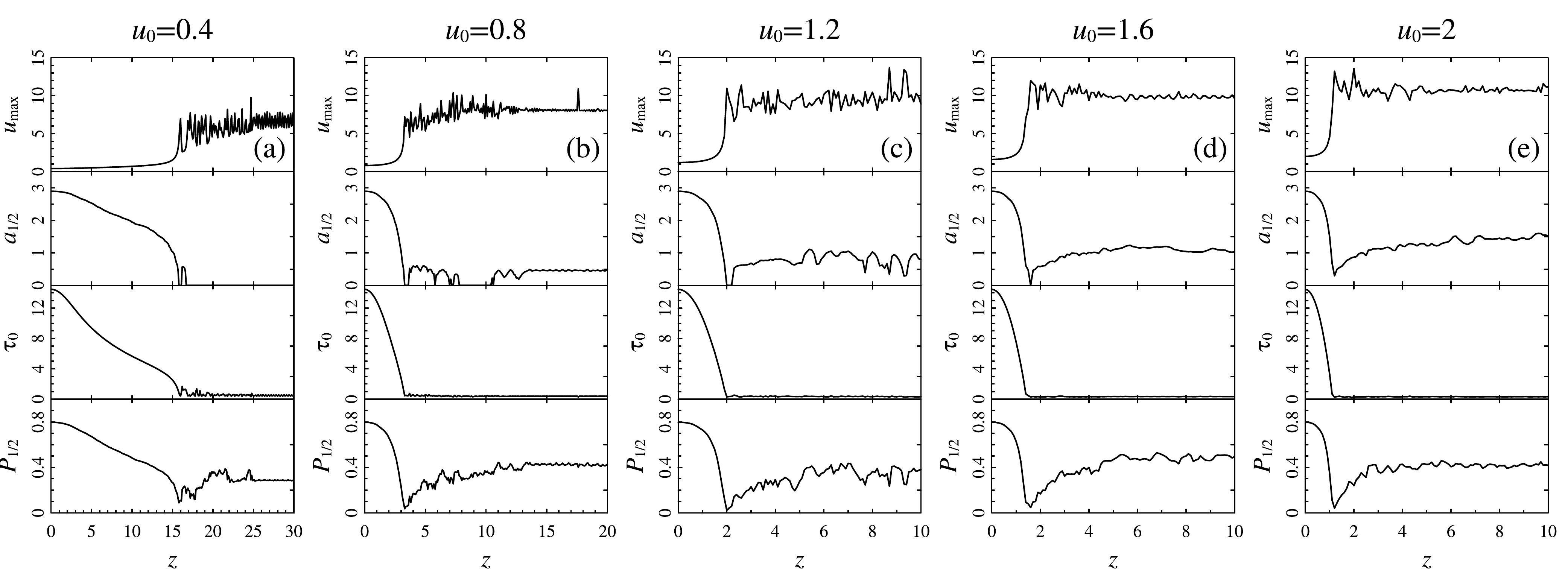}
	\caption{Maximal amplitude $u_\text{max}$, the width $a_\text{1/2}$, pike duration and the energy fraction $\eta_\text{1/2}$ in fibers with a high intensity depending on the evolution variable $z$ for the different initial field amplitude $u_0$. It is seen that limitation of the maximum amplitude is associated with the development of the modulation instability. In addition, one can see the lack of focusing in a single channel at the amplitude of $u_0 = 0.8$ and higher.}\label{ris:ris6}
\end{figure*}

Numerical simulation of the evolution of wave packets with the initial Gaussian form shows that the process of transverse self-focusing together with longitudinal self-compression develops in accordance with the variational approach analysis. The decrease in the pulse duration in the process of self-focusing (Fig.~\ref{ris:ris4}) occurs according to the law \eqref{eq:23}. So, we can obtain  from \eqref{eq:27} the following estimate for the self-focusing length:
\begin{equation}\label{eq:32_}
L_f=\dfrac{a_0^32\pi^{3/4}}{3W}.
\end{equation}
It is clear from Figs.\,\ref{ris:ris5} and \ref{ris:ris6}\textbf{(a)} that the light bullet is formed at this length, and then the radiation propagates in the self-channeling mode.

Self-focusing of the wave packet to a size of about the lattice period weakens diffraction effectively. Accordingly, mode \eqref{eq:23} of the adiabatic duration decrease becomes violated, and the pulse dynamics become close to the one-dimensional case. A strong increase in the field in the near-axis region creates conditions for the development of a modulation instability.
The situation here is the same as with the pulse propagation in a single fiber. One can see in Fig.~\ref{ris:ris5} the splitting of the wave packet into a set of soliton-type structures self-channeled in the central waveguide and the formation of three light bullets. 

Thus, the effect of the pulse compression in media with anomalous dispersion allows us to collect \emph{most} of the wave energy in a single light-guide having the pulse duration given by Eq.~\eqref{eq:32} (Fig.~\ref{ris:ris5} at $z=16$). However, the development of modulation instability limits the propagation of the fields with an amplitude being greater than the critical one in fiber systems. 

For larger pulse energies, the behavior of the characteristic parameters of the spatial part of the field (see Fig.~\ref{ris:ris6}) has a pattern which is similar in many aspects to that of the wave beam self-focusing (Fig.~\ref{ris:ris2}). As previously (Section~\ref{sec:4}), about half of the wave field energy (Fig.~\ref{ris:ris6}) is ejected into the peripheral region while passing from radiation self-focusing to self-channeling. At this, modulation instability occurs in each bright light-guide, which produces a set of light bullets with monotonically decreasing durations (Fig.~\ref{ris:ris6}) according to law \eqref{eq:23}.

\subsection{Hollow wave pulses}

Numerical simulation of the self-action shows that it is not possible to suppress the effects of modulation instability under conditions of enhanced filamentation with increasing energy in the pulse for the case of wave packets with the initially Gaussian shape. Forming a set of light bullets in the longitudinal direction is not a very good scenario. The task of summing the sequence of laser pulses at the output from a nonlinear medium for the purpose of further use is a complicated problem. It is a different matter in the case of predominant development of transverse filamentation and the subsequent formation of light bullets flying parallel to each other. Such an opportunity can, for example, be realized with axicon focusing in a system of coupled light guides \cite{Balakin}. This is due to the fact that in the Bessel wave beams the inhomogeneities of the transverse structure are pronounced much more clearly than in the Gaussian ones.

Let us consider the possibility of predominant development of the transverse filamentation of the wave field ahead of the longitudinal modulation in the problem of the plane wave instability. The generalization of the instability growth rate \eqref{eq:Gamma} to the case allowing for the additional weak longitudinal modulation with the scale $2\pi/\Omega$ leads to the following expression:
\begin{equation}\label{eq:33}
\Gamma^2=\left(\Omega^2+4\sin^2\frac{\kappa}{2}\right)\left(2 u_0^2-\Omega^2-4\sin^2\frac{\kappa}{2}\right) .
\end{equation}
Hence, discrete symmetry with respect to longitudinal and transverse perturbations is violated unlike in a continuous medium. The characteristic scales of the perturbations $2\pi/\Omega_0$, $2\pi/\kappa_0$, for which instability growth rate \eqref{eq:33} is maximal, are determined from the following relations:
\begin{subequations}\label{eq:34}
\begin{gather}
\sin\frac{\kappa_0}{2}\cos\frac{\kappa_0}{2}\left( u_0^2-\Omega_0^2-4\sin^2 \frac{\kappa_0}{2}\right)=0 \label{eq:34a} \\
\Omega_0^2+4\sin^2\frac{\kappa_0}{2}= u_0^2 \label{eq:34b} .
\end{gather}
\end{subequations}
As in Section \ref{sec:4}, it is natural to assume that the spatial inhomogeneity of the medium defines the characteristic spatial scale of the perturbations $\cos({\kappa_0}/{2}) = 0$ (see Eq.~\eqref{eq:Lperp}). In this case, we find for the longitudinal wave number from \eqref{eq:34b}
\begin{equation}\label{eq:35}
\Omega_0^2= u_0^2-4 .
\end{equation}
So, the development of the modulation instability is stabilized for the fields $ u_0<2$ and the transverse filamentation becomes dominant, which develops as in Section~\ref{sec:4}. The upper field limit ($ u_0<2$) is violated during the self-focusing and subsequent self-channeling of Gaussian wave beams (see Fig.~\ref{ris:ris6}). We suggest to use hollow wave packets to avoid pulse splitting in longitudinal direction.

\begin{figure*}
	\includegraphics[width=0.75\linewidth]{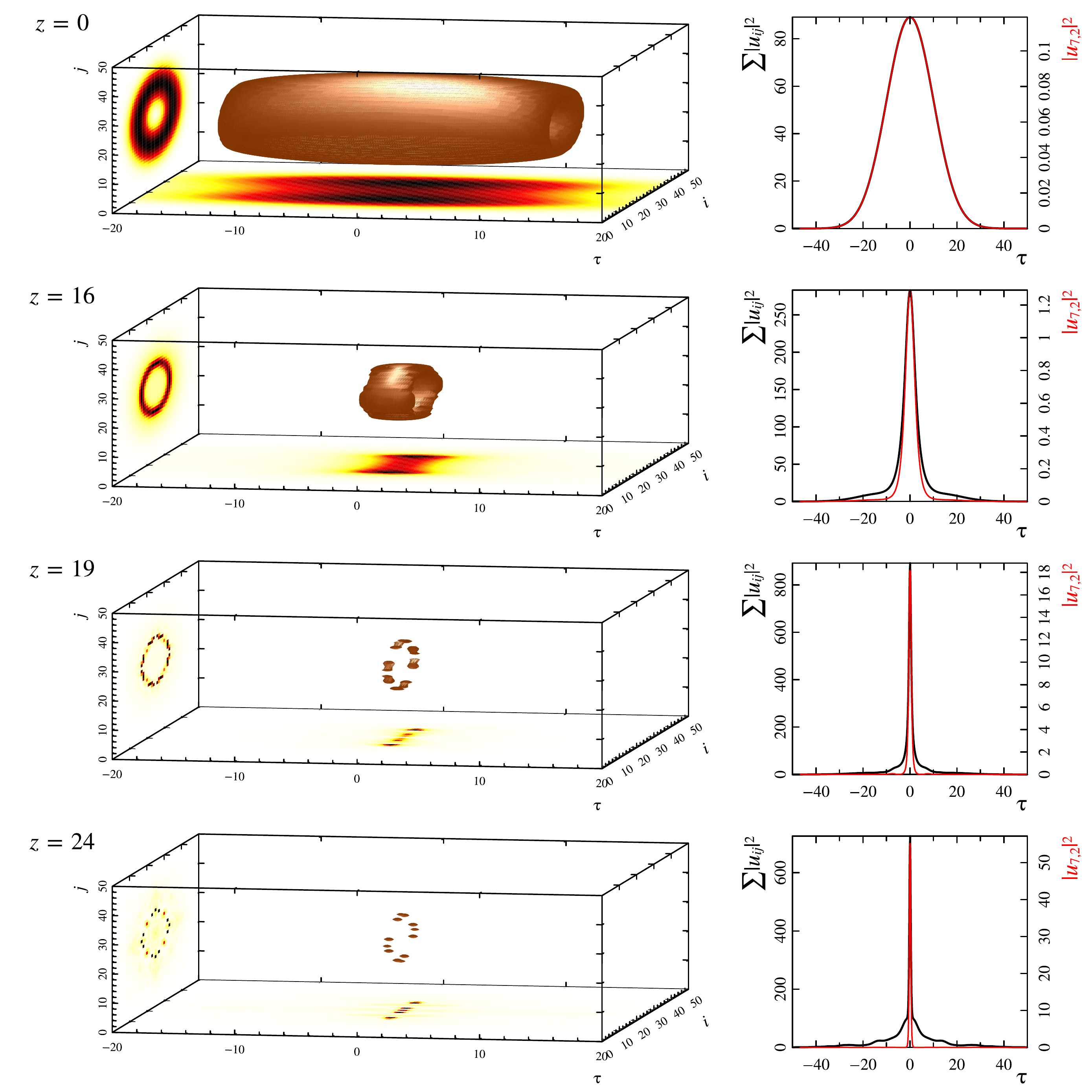}
	\caption{(Color online) The left column shows the amplitude distributions of the hollow wave packet at different values of $z$. The right column shows the corresponding time structures of the wave packet: integral one (black) and one in the region of the maximum field (red). The initial pulse is $u_{ij}=0.4 e^{-{\left(\sqrt{(i-25)^2+(j-25)^2}-10 \right)^2}/{50} - {\tau^2}/{100}}$.} \label{ris:ris7}
\end{figure*}

\begin{figure*}
	\includegraphics[width=0.72\linewidth]{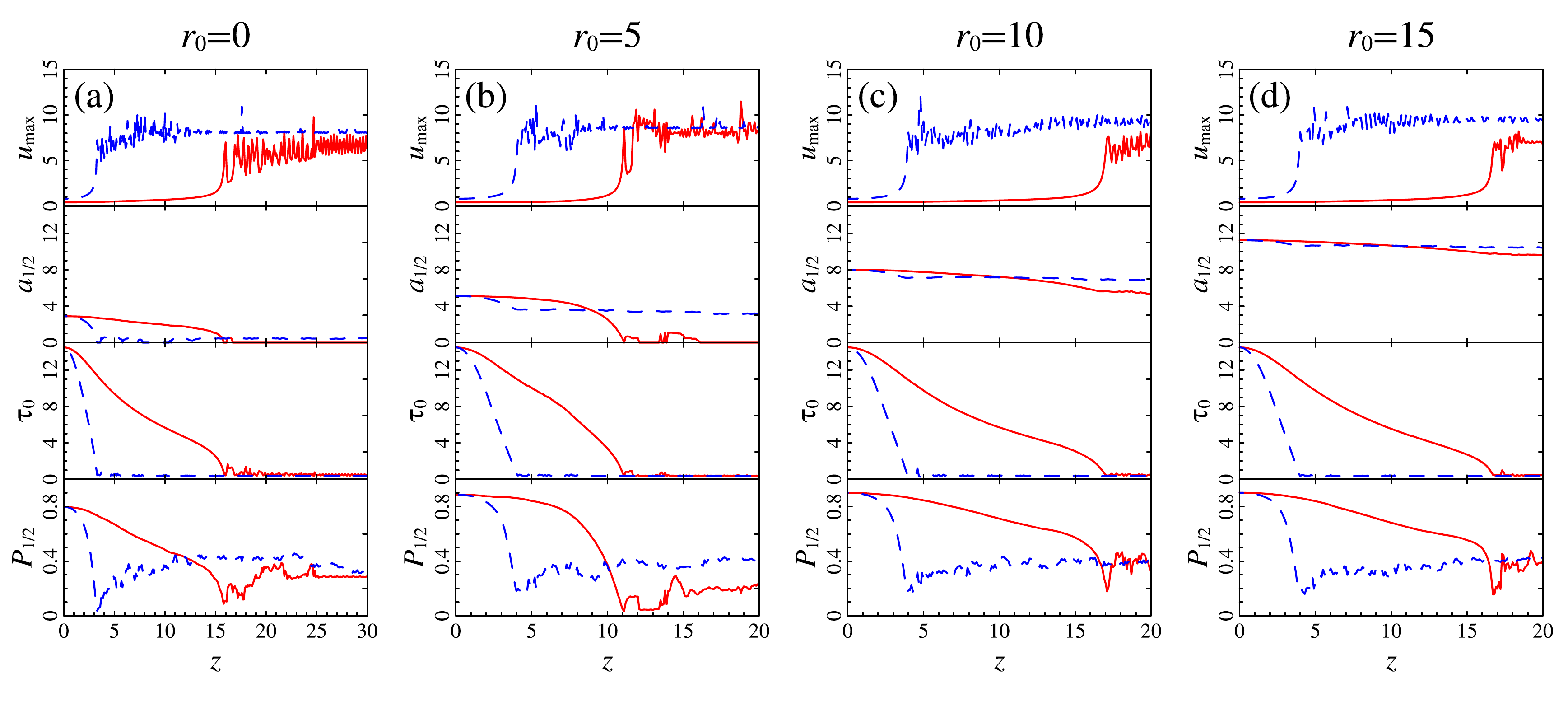}
	\caption{(Color online) Dependences of the maximal amplitude $u_\text{max}$, the width $a_\text{1/2}$, pike duration and the energy fraction $\eta_\text{1/2}$ in fibers with a high intensity depending on the evolution variable $z$ for different initial ring radius $r_0$ with the field amplitude $u_0=0.4$ (solid) and $u_0=0.8$ (dash). There is no collapse of the wave packet to the center fiber both for large initial amplitudes \textbf{(b)} and for a large radius of the ring \textbf{(c,d)}.}\label{ris:ris8}
\end{figure*}

The results of numerical simulation of the wave packets having the form
\begin{equation}\label{eq:36}
 u_{ij}=u_0 e^{-{\left(\sqrt{(i-25)^2+(j-25)^2}-r_0 \right)^2}/{50} - {\tau^2}/{100}}
\end{equation}
are shown in Figs.\,\ref{ris:ris7} and \ref{ris:ris8}. It is seen from Fig.~\ref{ris:ris8} that the self-focusing mode depends on the field amplitude. Thus, an increase in the field amplitude from $u_0 = 0.4$ to $u_0 = 0.8$ at $r_0 = 5$ leads to a change in the structure of the focal region. At $u_0 = 0.4$, the radiation self-focusing takes place on the axis of the system as in the continuous case. The hollow structure of the wave packet is preserved if the field amplitude is doubled. In this case, the self-focusing of the radiation is accompanied by an increase in the field on the ring, somewhat less than the initial radius (see Fig.~\ref{ris:ris8}), which weakens the maximum attainable value of the field amplitude. 
In the process of forming of an annular beam with a characteristic thickness being close to the lattice period, the laser pulse is shortened, and the field filamentation occurs only along the angular variable. The corresponding field evolution is shown in Fig.~\ref{ris:ris7}. One can see the formation of light bullets located on the ring and flying parallel to the axis of the system. The fifth part of the original energy is involved in this process.

\begin{figure}
		\includegraphics[width=\linewidth]{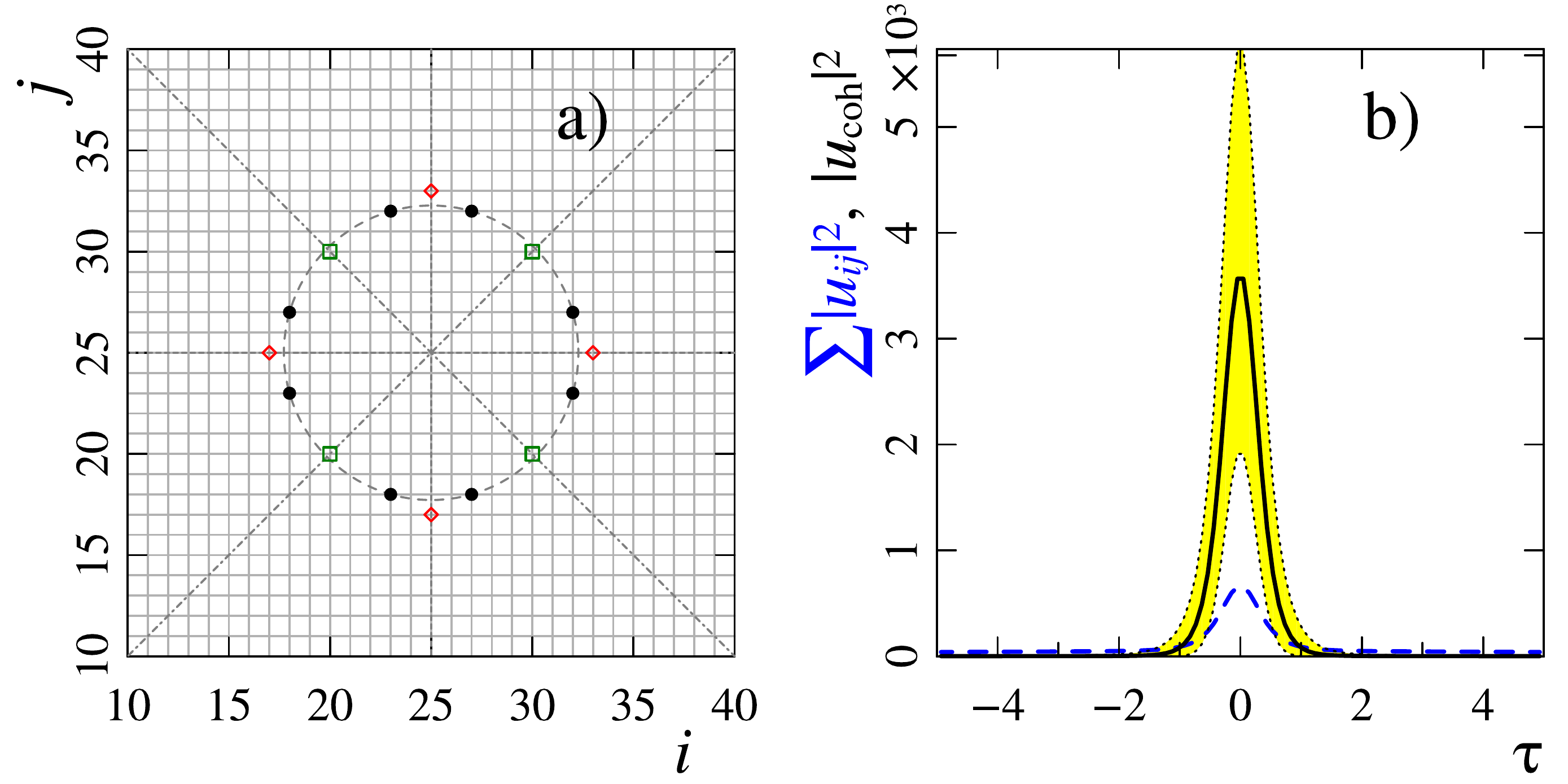}
		\caption{(Color online) Coherent groups of light-guides \textbf{(a)} and the result of coherent summation \textbf{(b)} of the wave field over most intensive light-guides (black solid circles in \textbf{(a)}) for the parameter in Fig.~\ref{ris:ris7}. The dashed line in the Figure \textbf{(b)} is the result of incoherent summation of the wave intensity over the whole region. The shaded area shows the change in the coherent intensity depending on the interaction length. } \label{ris:coh}
\end{figure}

As this  stratified structure propagates, the pulse duration decreases. The analysis of numerical simulation results show that such wave structures have equal phases only in certain groups of optical fibers located on the ring. This is due to the symmetry of the square structure of the array of fibers under consideration. Figure~\ref{ris:coh}\textbf{(a)} shows schematically three groups of light-guides (black solid circles, squares and diamonds), in which the wave structures are mutually in-phase. It should be noted that the fraction of energy contained in the group of optical fibers marked by black circles is noticeably larger than in the other groups. Figure~\ref{ris:coh}\textbf{(b)} shows the time distributions of the laser pulses for three different cases. The blue dash line shows the wave packet profile, which is the result of the summation $\sum |u_{ij}|^2$ of the intensities over all light guides. The solid black curve shows the time profile of coherent field from the group of light-guides, marked by black circles (see Fig.~\ref{ris:coh}\textbf{(a)}). It is important to note that the amplitude of the output signal does not depend on the interaction length after the formation of light bullets. 
Unfortunately, the wave field frequencies in different groups of light-guides are slightly different. This results in variation in the coherently summed amplitude of wave fields on the ring in dependence on the interaction length. The shaded area in Fig.~\ref{ris:coh}\textbf{(b)} shows the scatter of these profiles. It can be seen that the intensity of the coherent output signal in this case depends on the length of the light-guides. However, even in the worst case, the intensity of the coherent radiation is higher than the total ``incoherent'' intensity from all the light guides.
	
Thus, it is possible to gather the field from the group of light-guides marked by black circles using linear elements (for example, a lens) at the output of the system. This forms laser pulses which are much shorter and intense than the original one. As it follows from Fig.~\ref{ris:ris7} and \ref{ris:coh}\textbf{(b)}, the laser pulse duration decreases by a factor of 15 during the self-action process under consideration. Its intensity increases by more than 500 times. At this, the summed output pulse contains more than 25\% of the energy of the initial pulse.

\section{Conclusion}

The study of self-action of wave packets propagating along a two-dimensional system of coupled light-guides shows that media discreteness leads to new features in the evolution of the system. To construct a qualitative pattern, we developed the variational approach. This allowed us to classify the self-action regimes, determine the characteristic parameters, and describe the transition of self-focusing of wide wave beams to the self-channeling mode in the central fiber. The latter
%This process
reflects the peculiarities of a discrete medium and does not depend on the dimension of the problem.

A more detailed study of the wave field collapse and its subsequent self-channeling in the near-axis region was carried out on the basis of numerical simulation of the discrete NSE. The numerical analysis of self-focusing of wide axially symmetric wave beams has shown that the ``aberration-free'' self-action regime takes place in a limited range of powers $\mathcal{P} < 10 \mathcal{P}_\text{cr}$ exceeding the critical one for self-focusing. The process of the beam collapse is accompanied by ejection of about half of the wave field power from the near-axis region.

At powers exceeding the critical power of self-focusing by a factor of ten, the development of filamentation instability becomes a noticeable process. The characteristic scale of the instability is equal to the lattice period of the discrete medium and does not depend on the initial field amplitude. The maximal field amplitude is limited by value \eqref{eq:umax} due to breaking of the interaction between the neighboring light-guides for higher amplitudes. As a result, further propagation of the radiation occurs incoherently through several fibers in the self-trapping mode. This makes it impossible to collect a high-power wave beam into a single light-guide.

The analysis of the transition from laser pulse self-compression to its self-channeling shows the possibility of a noticeable decrease in the duration of a three-dimensional wave packet. However, as the self-focusing and localization of the radiation in the central light-guide increase, the field amplitude increases to such an extent that the variational approach becomes inapplicable due to the development of the modulation instability. The self-channeled compressed wave field splits into a set of solitons, as in the one-dimensional continuous case. Thus, the self-action of pulsed radiation in a discrete medium with an anomalous dispersion of the group velocity leads to formation of light bullets flying one-by-one along the system axis.

The formation of light bullets that fly parallel to each other is more convenient for practical use. Such situation can be realized under conditions when the filamentation instability develops faster than the modulation one. Hollow wave structures are suitable here. Self-focusing of such field distributions leads to a spatial localization of the wave packet near the ring. Numerical simulation shows that as a result of the development of filamentation instability along the angular variable, a set of light bullets is formed in the optical fibers located parallel to the axis of the system. In the process of self-action, the duration of the initial pulse also decreases. More importantly, solitons in individual fibers are coherent. Using linear elements (for example, a lens), the fields at the output of the system can be summed and form a much shorter  than the original, intense laser pulse.

This work was supported by the Russian Science Foundation (Project No. 16-12-10472).

\end{document}